\documentclass[12pt]{iopart}
\usepackage{iopams} 
\usepackage{graphicx}
\usepackage{subfigure}
\usepackage{xcolor}
\usepackage{soul}
 \expandafter\let\csname equation*\endcsname\relax 
 \expandafter\let\csname endequation*\endcsname\relax 
\usepackage{amsmath}
\usepackage{wasysym}
\usepackage{amssymb}
\usepackage{color}
\usepackage{soul}
\usepackage{ulem}

\begin{document}

\title[Preferred degree networks]{Modeling interacting dynamic networks: I. Preferred degree networks
and their characteristics}

\author{Wenjia Liu$^{1}$, Shivakumar Jolad$^{3}$, Beate Schmittmann$^{1,2}$, and R. K. P. Zia$^{1,2}$}

\address{$^1$Department of Physics and Astronomy, Iowa State University, Ames, IA 50011, USA}
\address{$^2$Department of Physics, Virginia Tech, Blacksburg, VA 24061}
\address{$^3$Indian Institute of Technology - Gandhinagar, Ahmedabad, Gujarat 382424, India}

\ead{wjliu@iastate.edu, shiva.jolad@iitgn.ac.in, schmittb@iastate.edu and rkpzia@vt.edu}
\begin{abstract}
We study a simple model of dynamic networks, characterized by a set preferred degree, $\kappa$. Each node with degree $k$ attempts to maintain its $\kappa$ and will add (cut) a link with probability $w(k;\kappa)$ ($1-w(k;\kappa)$). As a starting point, we consider a homogeneous population, where each node has the same $\kappa$, and examine several forms of $w(k;\kappa)$, inspired by Fermi-Dirac functions. Using Monte Carlo simulations, we find the degree distribution in steady state. In contrast to the well-known Erd\H{o}s-R\'{e}nyi network, our degree distribution is not a Poisson distribution; yet its behavior can be understood by an approximate theory. Next, we introduce a second preferred degree network and couple it to the first by establishing a controllable fraction of inter-group links. For this model, we find both understandable and puzzling features. Generalizing the prediction for the homogeneous population, we are able to explain the total degree distributions well, but not the intra- or inter-group degree distributions. When monitoring the total number of inter-group links, $X$, we find very surprising behavior. $X$ explores almost the full range between its maximum and minimum allowed values, resulting in a flat steady-state distribution, reminiscent of a simple random walk confined between two walls. Both simulation results and analytic approaches will be discussed.

\end{abstract}

\maketitle

\section{INTRODUCTION}

Networks are ubiquitous, emerging in natural structures as well as man-made artifacts. Examples of the former range from the microsopic, e.g., neuron
architectures, to the cosmic, e.g., galactic filaments. For the latter, they include critical infrastructures, e.g., power or transportation grids, as well as virtual webs, such as Facebook and LinkedIn. Understanding their characteristics and
behaviors is clearly important \cite%
{Strogatz01,AlbertBarabasi02,DorogovtsevMendes02,Newman03,EstradaFoxHighamOppo10}%
. In recent years, the statistical properties of complex networks such as
their topology \cite{WattStrogatz98,AlbertJeongBarabasi99}, their evolution
over time \cite{DorogovtsevMendes02,BarabasiAlbert99}, and dynamical
processes on them \cite{GrossBlasius08,BarratBarthelemyVespignani08} have
been widely investigated. However, most of these studies have focused on
single isolated networks. By contrast, networks in the real
world are often intimately intertwined. At the cellular level, drastically
different networks, from the cytoskeleton to regulation and signaling, form
intricate patterns of interdependence. Similarly, infrastructure networks such as airlines, ground transportation, power grids, and telecommunications are highly interdependent. Meanwhile, the internet
plays a critical role by interacting with all of them. In the last few
years, the significance of \textit{interacting} networks is coming onto
center-stage, and many scientists and engineers turn their attention to various aspects
of such interactions. Examples of these studies include critical 
infrastructure interdependencies {\cite%
{RinaldiPeerenboomKelly01,PanzieriSetola08,BuldyrevParshaniPaulStanleyHavlin10,BuldyrevShereCwilich11}, and approaches such as  the multilayer method for modeling traffic flows on an underlying infrastructure \cite{KurantThiran06}. 

Since real interacting networks are extremely complex, even building good models for them is already challenging, not to mention developing reliable analytic approaches. Our goal here is to introduce a model that is sufficiently simple so that 
analytic solutions are within our reach. Further, in deference to realistic interacting networks, our links will be \textit{%
dynamic}. Specifically, our model is motivated by social networks in which
the nodes are individuals and the links represent contacts between them. As social connections are made and broken, links are added or cut, resulting in a dynamically evolving network structure. In the long term, the goal is to
investigate not only the interactions between such dynamic networks, but also to include the degrees of freedom associated with the nodes, e.g., opionion, wealth, health, etc. \cite%
{GrossBlasius08,BarratBarthelemyVespignani08,JoladLiuSchmittmannZia12,ZanetteR.JBioPhys.2008,GrossK.EPL.2008,ShawS.PRE.2008,SchwarzkopfRM.PRE.2010,MarceauPHAD.PRE.2010,WangCSA.JPhysA.2011}.
In such models, the
network structure and the attributes of the nodes co-evolve, coupled by nontrivial dynamic feedback. 

In a series of papers, we present a model of dynamical links which we
believe reflects natural and typical human behavior, i.e., a network with 
\textit{preferred degrees}. In any society, individuals have their preference for associating with a certain number of friends. We model this preference by a `preferred degree' ($\kappa$) for each node. Of course, $
\kappa $ depends on the `personality' of the individual and varies from
person to person. Further, it may change intrinsically with time; perhaps 
slowly as one ages and the need for contacts changes. 
$\kappa $ may also adapt to
varying external circumstances, such as moving to a different neighborhood, joining a workplace or escaping a raging epidemic or a war. 
Many authors have tried to model such dynamic networks by introducing a `rewiring' of links \cite{GrossBlasius08,BarratBarthelemyVespignani08}. While this has the (technical) advantage of keeping the total number of links constant, in some circumstances, it might be more realistic to remove this constraint. For example, in the 2003 SARS epidemic outbreak \cite{http://www.moe.gov.sg/media/press/2003/pr20030326.htm}, the Singapore government closed all schools which drastically cut down interactions among children. Obviously, the students did not
`rewire' to $O\left( 100\right) $ new contacts in order to replace the
classmates lost by this decree. Instead, such a situation can be modeled by a $\kappa $ which drops by an order of magnitude (e.g., from 100 classmates to
10 family members). A more detailed discussion of epidemic spreading on preferred degree networks is beyond the scope of this paper but can be found in \cite{JoladLiuSchmittmannZia12}. 

In this first paper of a series, we focus exclusively on the dynamics of links, i.e., the number of nodes and their degree $\kappa $ remain fixed, and the nodes carry no degrees of freedom.
Since interacting networks are complex, we will begin with a `baseline
study,' namely, the simplest possible preferred degree network -- a
homogeneous population -- in which all individuals have the same $\kappa $
and do not discriminate against, or in favor of, any other individual when
adding or cutting links. We will introduce rules on how an individual node decides
to add or cut a link and discuss the degree distribution after the system
has settled into a stationary state: $\rho ^{ss}\left( k\right) $. Despite
the appearance of randomness (with mean degree $\kappa $), our steady-state degree distribution $\rho
^{ss}\left( k\right) $ differs significantly from the Poisson form found in the standard
Erd\H{o}s-R\'{e}nyi random network \cite{E-R}.

Then, we will introduce two such networks and couple them with arguably the
simplest of couplings: When an individual adds or cuts a link, this action is performed on a inter-network link with fixed probability $\chi $. Such
a deceptively simple generalization leads to remarkably rich behaviors, as we
extend the characterization to distributions of \textit{intra-} and \textit{%
inter-}network degrees. In particular, we begin with networks which differ
by only one of the three parameters, $\left( N,\kappa ,\chi \right) $, where 
$N$ is the number of nodes. We find that the total number of crosslinks
between the two networks, $X$, plays a key role in characterizing the
interactions, while its stationary distribution, $P^{ss}\left( X\right) $,
displays highly non-trivial behavior.

This paper is organized as follows: In section $2$, we present the
specifications of our model, first for a homogeneous population with a single
preferred degree and then a heterogeneous one consisting of two different groups. We also introduce several quantities that serve to characterize the
interaction between the two groups, e.g., the distribution of crosslinks. In
section $3$, we present the results from Monte Carlo simulations along with an approximate mean field theory which allows us to understand some of the remarkable behaviors. We close with a summary and outlook in
section $4$, while deferring some technical details to the Appendices.

\section{SPECIFICATIONS OF THE MODELS}

Since our goal is the behavior of interacting networks with preferred
degrees, we begin with a detailed description of a network with a single
preference, namely a homogeneous population. Thereafter, we will describe a simple way to induce interactions between such networks and study their properties. Such interacting networks can be considered as a highly idealized model of a society with inhomogeneous populations.

\subsection{Homogeneous populations}

Though the notion of a preferred degree network has been introduced
previously  by the present authors and their collaborators \cite{PlatiniZia10,ZiaLiuJoladSchmitt11,JoladLiuSchmittmannZia12,LiuSchmittmannZia12}
, let us recapitulate the main ingredients here. We consider a network
consisting of $N$ (static) nodes (with no degrees of freedom) and a dynamically
evolving set of links. Each node is endowed with the same fixed attributes: $
w_{+}\left( k\right) $ and $w_{-}\left( k\right) $, the probability that it
will create and destroy a link, respectively, given that it has $k$ links. For simplicity, we restrict our study to the special case $
w_{-}=1-w_{+}$ and denote $w_{+}\left( k\right) $ by $w\left( k\right) $. Starting from an empty network, we generate links as follows. In each time step (attempt), we select a node at random and find its degree, $k$. Then, with probability $w(k)$, this node creates a new link to a randomly chosen node, which is not already connected to it. Otherwise, this node cuts one of its existing links at random, with probability $1-w(k)$. Self-loops and multiple connections are not allowed. To model an individual's natural preference for some finite number of contacts, 
$\kappa $, we choose a $w$ to be close to $1$ when $k\ll \kappa $, crossing
over to $\thicksim 0$ when $k\gg \kappa $. The simplest $w$ with this
property is the step function {$\Theta (\kappa -k)$ (i.e., $1$ for }$k\leq
\kappa $ and $0$ otherwise), as a good model for a very `rigid personality,'
adding or cutting links as soon as the preferred degree is unmet or
exceeded. We have also considered more moderate `personalities,' modeled by a Fermi-Dirac function: 
\begin{equation}
w(k)=\frac{1+e^{-\beta \kappa }}{1+e^{\beta (k-\kappa )}}  \label{w}
\end{equation}%
where $\beta $ plays the role of an individual's `rigidity.' Although there are innumerable ways to choose the form of $w(k)$, we restrict our analysis to the form in  Eq.~(\ref{w}). Studying the details of $w(k)$ is not the primary goal here. Instead, our interest focuses on universal statistical
properties of the collective behavior of many individuals, such as the
degree distribution, which we denote by $\rho \left( k\right)$.

\subsection{Heterogeneous populations with two groups}
\begin{figure} 
\centering
\includegraphics[width=3in]{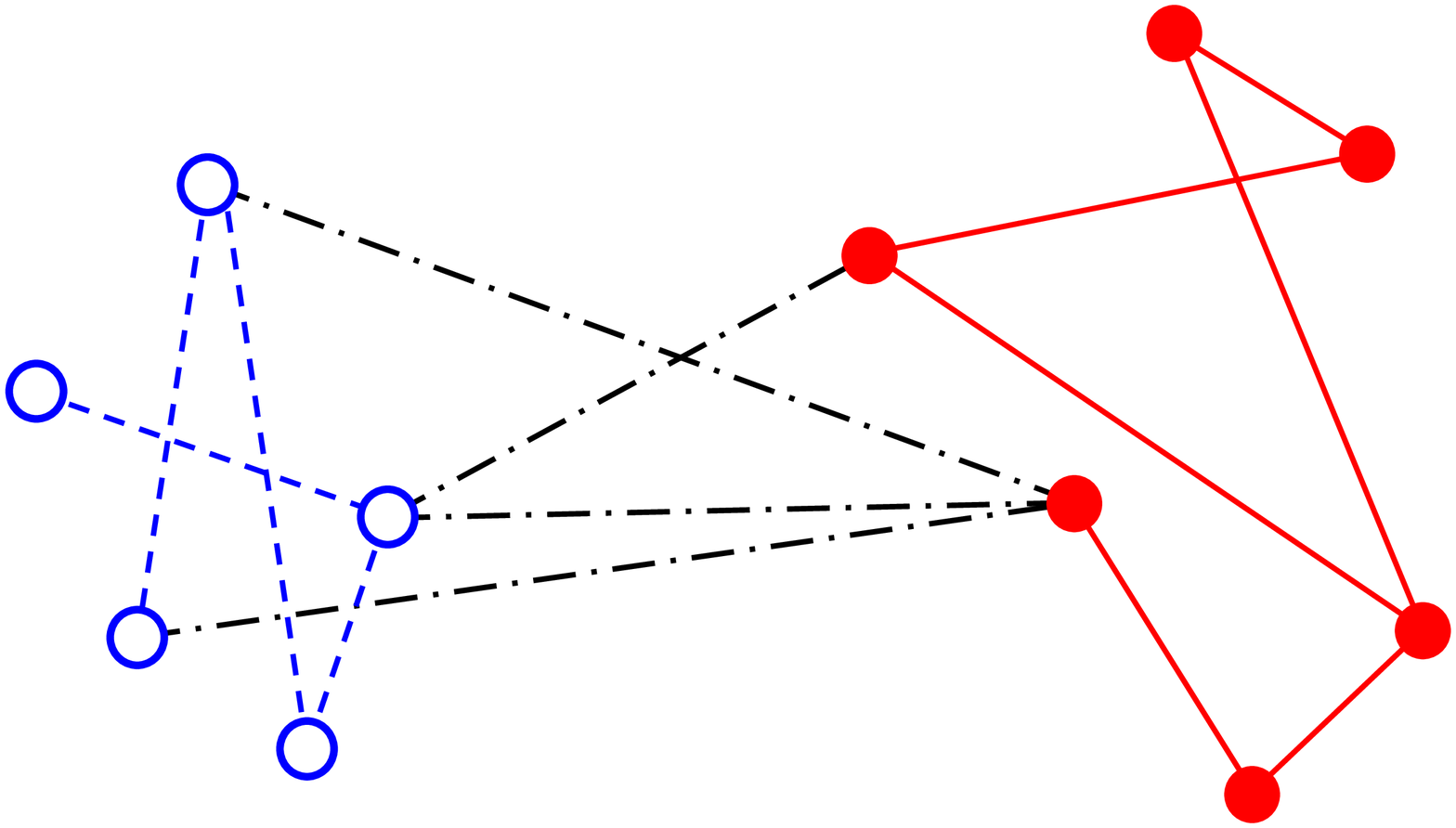}%
\caption{\label{2} The nodes of the two groups are denoted by open blue ($1$) and closed red ($2$)
circles. The intra-group links are shown as blue dashed and 
red solid lines, while the inter-group links are
dot-dashed lines (black). For this network, the sets of $k$'s are: 
$k_{1}^{\ast }=\left\{ 2,1,2,1,2\right\} $, $k_{1}^{\times }=\left\{
1,0,2,1,0\right\} $, $k_{2}^{\times }=\left\{ 0,0,1,3,0,0\right\} $, and 
$k_{2}^{\ast }=\left\{ 2,2,2,1,3,2\right\} $. Thus, the non-vanishing
contributions to the distributions are $\rho _{1}^{\ast }\left( 1\right)
=2,\rho _{1}^{\ast }\left( 2\right) =3,\rho _{1}^{\times }\left( 1\right)
=2,\rho _{1}^{\times }\left( 2\right) =1$ and $\rho _{2}^{\times }\left(
0\right) =4,\rho _{2}^{\times }\left( 1\right) =1,\rho _{2}^{\times }\left(
3\right) =1,\rho _{2}^{\ast }\left( 1\right) =1,\rho _{2}^{\ast }\left(
2\right) =4,\rho _{2}^{\ast }\left( 3\right) =1$.}
\label{nodes}
\end{figure}

Keeping in mind that our distant goal is the interaction between networks
of possibly distinct types, our next step in that direction is modest: the
study of two networks of the same type, namely, two preferred degree networks. Needless to say,
in a typical heterogeneous society, individuals will have a range of
preferences and flexibilities. As a first attempt, we will focus on a
population with just two different preferences ($\kappa _{1}<\kappa _{2}$),
but the same personality (rigid). In particular, we will refer to the first
group as `introverts,' since these individuals prefer fewer contacts, while
group $2$ consists of `extroverts.' We will also allow the groups to have
different sizes in general, with {$N_{1}\neq $ $N_{2}$ nodes, i.e., a total
of $N\equiv N_{1}+N_{2}$ nodes.}

If there is no contact between two such groups, the above forms a complete
set of specifications. But, as soon as we wish to model their interactions,
many possibilities arise. In this first study, we restrict
ourselves to only one mechanism, by introducing a new parameter $\chi \in \lbrack 0,1]$, the probability that a node interacts with an inter-group node. In an attempt, a random node from the entire population is chosen. Given the chosen node's degree, $k$, $w(k)$ will determine if a link is to be added or cut. Whether adding or cutting, the action
will be executed on an \textit{inter}- or an \textit{intra}-group link with probability $\chi$ or $1-\chi$. In other words, 
$\chi $ is the chance that the action (adding or cutting) is taken on a 
\textit{crosslink}.} In general, the two groups can be assigned different
values: {$\chi _{1}\neq $ }$\chi ${$_{2}$}. In a sense, these are associated
with how strongly the networks interact, since {$\chi _{1}=$ }$\chi ${$_{2}$}%
$=0$ corresponds to a system consisting of two independent, homogeneous networks. At
the other extreme, {$\chi _{1}=$ }$\chi ${$_{2}=$}$1$ models a system with
crosslinks \textit{only}. Meanwhile, it is natural to expect the most
`symmetric' case, {$N_{1}=N_{2},\kappa _{1}=\kappa _{2},{\chi _{1}=}\chi _{2}=1/2$} to correspond to a single, homogeneous population. Yet,
surprising differences emerge when simulations are carried out.

Given that we are dealing with two networks, a variety of degree distributions can be defined. Specifically, in addition to $\rho
_{\alpha }$, the distribution of the total number of contacts associated
with a node in group $\alpha $ ($\alpha =1,2$ in this study), we will study
four other degree distributions, namely, the \textit{intra}- and \textit{inter}-group degree distributions of each network. Let us introduce the notations in the following. Let $k_{\alpha }^{\ast }$ and 
$k_{\alpha }^{\times }$ be associated with, respectively, the \textit{intra}%
-group and \textit{inter-}group degrees of a node in group $\alpha $. A
specific example is provided in Figure 1, with 
($k_{1}^{\ast},k_{1}^{\times },k_{2}^{\ast },k_{2}^{\times }$) specified explicitly in the caption. In
this way, for each group, we will study two distributions, $\rho _{\alpha
}^{\ast }\left( k_{\alpha }^{\ast }\right) $ and $\rho _{\alpha }^{\times
}\left( k_{\alpha }^{\times }\right) $. With this notation, it is clear that
generically, $\rho _{1}^{\times }\neq \rho _{2}^{\times }$. We should
emphasize that, though the `total degree' is just the sum ($k_{\alpha
}=k_{\alpha }^{\ast }+k_{\alpha }^{\times }$), the distributions $\rho
_{\alpha }$, $\rho _{\alpha }^{\ast }$, and $\rho _{\alpha }^{\times }$
contain very different information. For example, in the next section we will demonstrate that $\rho _{\alpha }^{\ast }$ and $\rho _{\alpha }^{\times }$ are
typically Gaussian-like, while $\rho _{1,2}$ are simple two-tailed exponential
distributions.

A more global quantity of interest is $X$, the \textit{total} number of
crosslinks in the system. It is natural to consider its distribution, which
we denote by $P\left( X\right) $. Of course, there is no \textit{a priori}
reason to expect $P$ and the $\rho $'s to be related in a simple way. Indeed, we will find that they are generally quite different. Nevertheless,
when the system settles in a stationary (steady) state, their averages must
obey the equalities%
\begin{equation}
\left\langle X\right\rangle =N_{1}\left\langle k_{1}^{\times }\right\rangle
=N_{2}\left\langle k_{2}^{\times }\right\rangle
\end{equation}%
where $\left\langle X\right\rangle \equiv \sum_{X}XP^{ss}\left( X\right) $
and $P^{ss}$ is the steady-state distribution, etc. In this paper, we will show that $P^{ss}\left( X\right) $ is a rather unusual distribution. In particular, even for populations of 
$O\left( 10^{3}\right) $, we observe that  
$X\left( t\right) $ wanders so slowly that its distribution cannot be reliably measured. Instead,
only by restricting ourselves to systems of $O\left( 100\right) $ are we
able to find $P^{ss}\left( X\right) $ with confidence. In the next section,
more details and our findings will be presented.

\section{SIMULATION RESULTS AND THEORETICAL CONSIDERATIONS}

With the models clearly specified, we can compute all quantities of
interest, \textit{in principle}, and predict the behavior of this system. In
practice, the mathematical challenges are insurmountable and, to gain
insight into the statistical properties, we perform Monte Carlo simulations
on the one hand and on the other, formulate approximation schemes which can
capture the main features.

\subsection{Statistical properties of a single network}
\begin{figure*} 
  \centering
  \mbox{
    \subfigure{\includegraphics[width=3in]{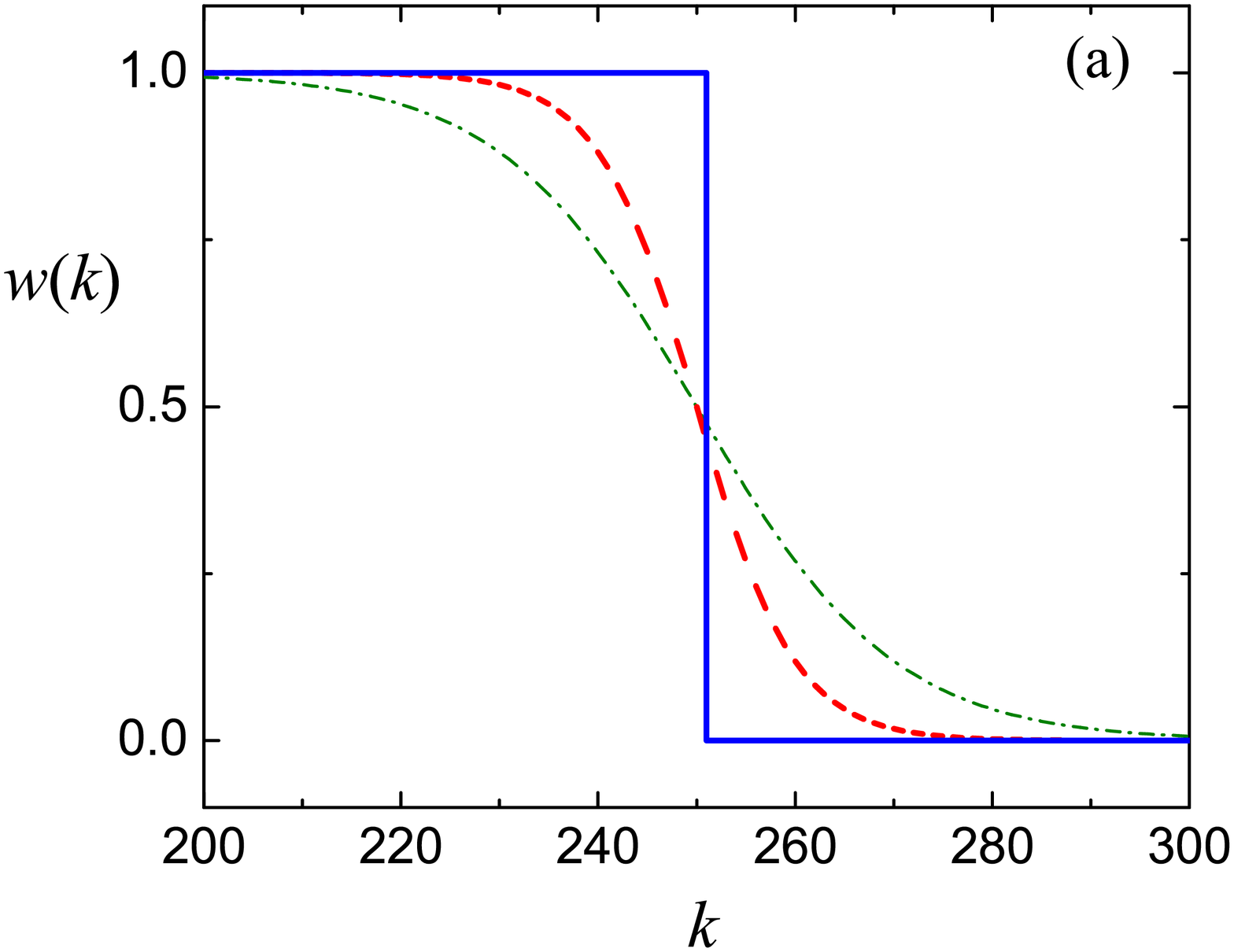}}\quad
    \subfigure{\includegraphics[width=3in]{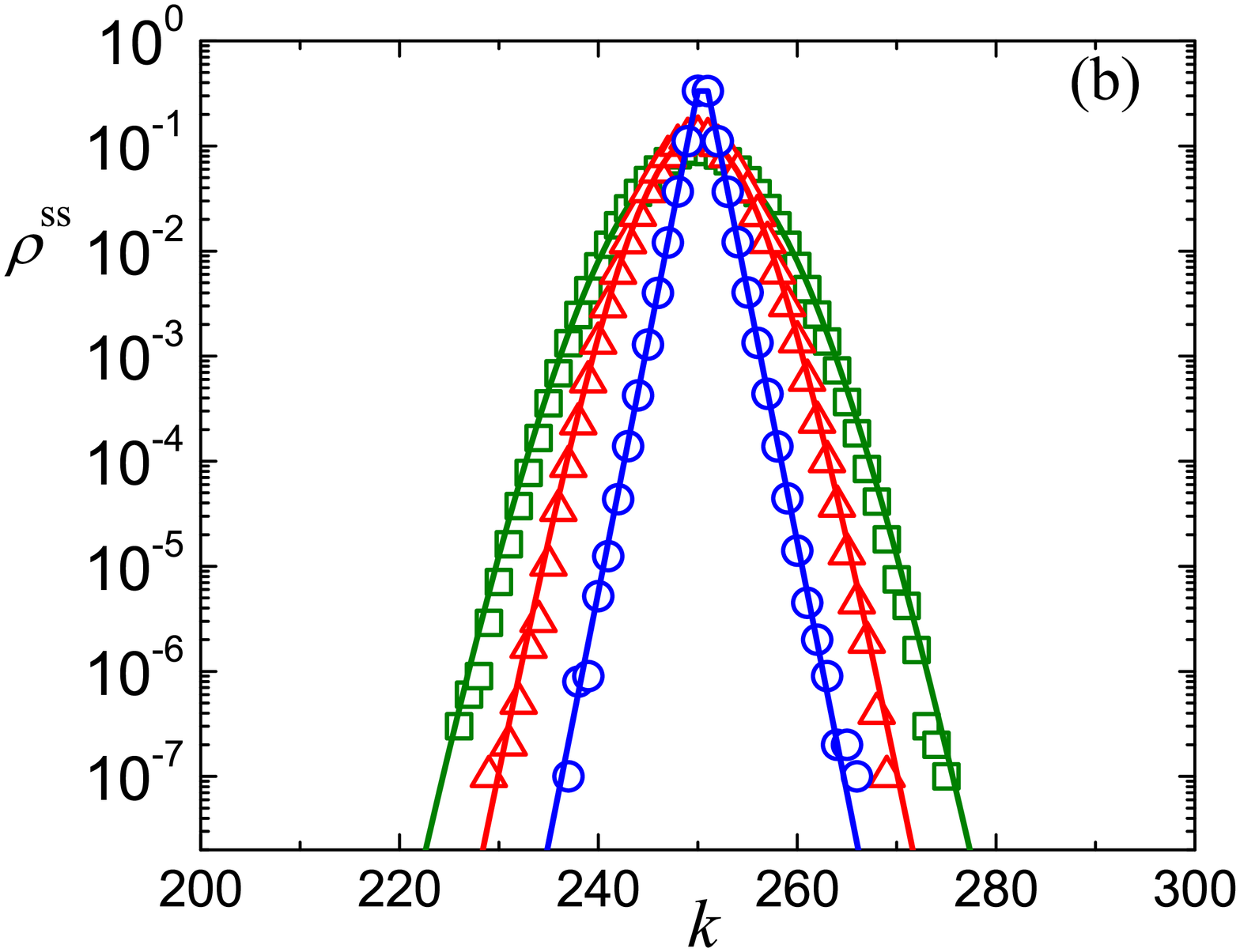}}
     }
  \caption{(a) Three different Fermi functions $w(k)=(1+e^{-\beta\kappa})/(1+e^{\beta(k-\kappa)})$, for $\kappa=250$: The green dash-dotted line, red dashed line and blue solid line represent $\beta=0.1$, $0.2$ and $\infty$ respectively. (b) The data points represent the corresponding degree distributions of a system  with $N=1000$. The solid lines are theoretical predictions.}
\label{single}
\end{figure*}

We first establish a baseline for our study, by investigating a homogeneous population. We choose reasonably large $N$ ($1000$)
and $\kappa $ ($250$), as well as three different $w$'s ($\beta
=0.1,0.2,\infty $, shown in Fig.~\ref{single}(a)). To facilitate comparisons
between the different cases of $\beta $, we actually use half integer $%
\kappa $'s, e.g., $250.5$. Starting with an empty network, we follow the
stochastic rules above and generate a new configuration with each attempt
(at adding/cutting links). Defining a Monte Carlo step (MCS) as $N$ updates,
we ensure that each node has one chance, on the average, to update its links
in a MCS. We discard the first $1K$ MCS, which appears to be sufficient here
for the system to reach steady state. Thereafter, we measure the quantities
of interest every $100$ MCS and compile averages from $10^{4}$ measurements
(i.e., runs of $1M$ MCS). Denoting the number of nodes with $k$ links in
each measurement by $n_{k}$, we find $\rho \left( k\right) $ through%
\begin{equation}
\rho (k)=\frac{\langle n_{k}\rangle }{N},
\end{equation}%
as illustrated in Fig.~\ref{single}(b).

For the well known Erd\H{o}s-R\'{e}nyi random network, $\rho^{ss}(k)$ is a
Poisson distribution with mean being the average degree. In our network, we clearly
expect $\left\langle k\right\rangle \cong \kappa $, and, given that links
are created and destroyed at random, we may also expect a Poisson
distribution. However, as illustrated in Fig.~\ref{single}(b), our
simulations show otherwise. In the simplest case ($w$ being a
step function), $\rho^{ss} (k)$ is consistent with a two-tailed exponential distribution, $%
\propto e^{-\mu \left\vert k-\kappa \right\vert }$ (blue circles in Fig.~\ref{single}(b)).
For less rigid populations, $\rho^{ss} (k)$ {depends on the details of $w(k)$ for 
$k\thicksim \kappa $, (green squares and red triangles) but
crosses over to the same exponential tails. Our data indicate $\mu =1.08\pm 0.01$. 
In the next few paragraphs, we will present an approximate theory,
shown as solid lines in the figure, which provides excellent agreement with
this data.

The full description of the stochastic dynamics of our network requires
writing down and solving the master equation for the probability of each
configuration. Since the mathematical details are quite involved, we present the main results here and leave the technicalities  to Appendix A. In particular, we find that
the dynamics violates detailed balance so that it is essentially impossible
to solve for the exact stationary probability distribution, let alone to
compute a quantity like $\rho \left( k\right) $. Thus, finding a reliable
approximation scheme is crucial for progress. One possibility is to
postulate an equation for $\rho \left( k,t\right) $ directly and compare its
predictions with simulation data. Approximating the evolution of $\rho
\left( k,t\right) $ by a Markovian birth-death process, we write an expression for $\rho
\left( k,t+1\right) -\rho \left( k,t\right) $:%
\begin{eqnarray}
&&W\left[ k,k+1\right] \rho \left( k+1,t\right) -W\left[ k+1,k\right] \rho
\left( k,t\right)  \label{1} \\
&&~~-\left\{ W\left[ k-1,k\right] \rho \left( k,t\right) -W\left[ k,k-1%
\right] \rho \left( k-1,t\right) \right\}  \label{2}
\end{eqnarray}%
where $W[k,k^{\prime }]$ specifies the rate for a node with degree $%
k^{\prime }$ to change to $k$. Since $k$ is non-negative, 
Eq.~(\ref{2}) is
absent for the $k=0$ case. Note that we have cast this expression as the
difference of two (probability) currents: Eq.~(\ref{1}) being the net current
from $k+1$ to $k$ and (\ref{2}), from $k$ to $k-1$. The advantage of this
form is that, in the stationary state, all these currents must vanish,
leading to%
\begin{equation}
W[k-1,k]\rho ^{ss}(k)=W[k,k-1]\rho ^{ss}(k-1).  \label{ME-rho}
\end{equation}%
Thus, once the $W$'s are given, the stationary $\rho ^{ss}(k)$ can be found
explicitly.

Next, we approximate $W$ by the following arguments. Focusing on a
particular node $i$, we note that the contributions to $W[k-1,k]$ come
from two processes. In one process, node $i$ is chosen and a link is cut with
probability $1-w\left( k_{i}\right) $. In the second process, one of the
other nodes, $j$, connected to $i$ is chosen and cuts its link to $i$. To account
for this rate exactly is quite involved, since there are $k_{i}$ such nodes,
all having varying degrees. Further, once $j$ is selected, the probability that it cuts its link to $i$ is $1/k_{j}$. Thus, we propose the
following rough estimate. We expect that, in a steady state, half of the
nodes have \textquotedblleft too many\textquotedblright\ links and, when
chosen, will cut. This provides a factor $k_{i}/2$, which we approximate by $%
\kappa /2$. Meanwhile, we approximate the various $1/k_{j}$'s by $1/\kappa $%
. The result is that the probability of this second process is simply $1/2$,
so that%
\begin{equation}
W[k_{i}-1,k_{i}]\thicksim \left\{ 1-w\left( k_{i}\right) +1/2\right\} /N
\label{W}
\end{equation}%
A slightly more sophisticated scheme, along with a more carefully detailed
argument, can be found in Appendix B. Since the results are not
significantly different, especially in cases with moderate $\kappa $'s, we
will continue to rely only on the simple picture here.

A similar argument leads to $W[k_{i},k_{i}-1]\thicksim \left\{ w\left(
k_{i}-1\right) +1/2\right\} /N$, so that we obtain%
\begin{equation}
\frac{\rho ^{ss}(k)}{\rho ^{ss}(k-1)}=\frac{w(k-1)+1/2}{1-w(k)+1/2}
\label{rho^ss}
\end{equation}%
in this approximation. For `rigid' personalities, the solution is trivial, 
\begin{equation}
\rho _{{}}^{ss}(k;\beta =\infty )\propto 3^{-\left\vert k-\kappa \right\vert
}  \label{rho^ss-rigid}
\end{equation}%
In other words, this crude scheme predicts $\mu =\ln 3\cong 1.0986$, a value
remarkably close to the one observed. For more `flexible' individuals, the $%
w $'s vary gently around $\kappa $, but cross over to $0$ or $1$ for $k\gg
\kappa $ or $k\ll \kappa $. Thus, the kink in $\rho ^{ss}\left( k\right) $
for $k\cong \kappa $ softens, crossing over to tails governed by the same
exponential. For the two cases with finite $\beta $ we simulated, we solve
Eq.~(\ref{rho^ss}) numerically and show the resultant as solid curves in
Fig.~\ref{single}(b) \cite{JoladLiuSchmittmannZia12}. Clearly, despite its crudeness, this approximation
captures the essense of the steady-state degree distribution.

\subsection{Interacting networks with different characteristics}

With the statistical properties of a single preferred degree network reasonably
well understood, we turn to a system with just two networks and focus on
the effects of coupling them in a particular manner (i.e., through $\chi >0$
described in Section II). Even with this restriction, the varieties of
having two networks are limitless. Thus, we further restrict ourselves to
studying only `rigid' individuals ($\beta =\infty $), and groups which are
identical except for {$N_{1}\neq N_{2},\kappa _{1}\leq \kappa _{2}$, and, in
some cases, ${\chi _{1}}\neq \chi {_{2}}$}. In other words, we attempt to
model the interactions between `rigid' introverts and extroverts, albeit in a very simplified fashion. 
To simulate the model described above, one MCS is defined as $N
=N_{1}+N_{2}$ updates. Thus, each node is again given one chance, on the average, to
take action. 

\subsubsection{Equal $N$'s and $\protect\kappa $'s, but {${\protect\chi _{1}}%
\neq \protect\chi {_{2}}$}.}
\begin{figure*} 
  \centering
  \mbox{
    \subfigure{\includegraphics[width=3in]{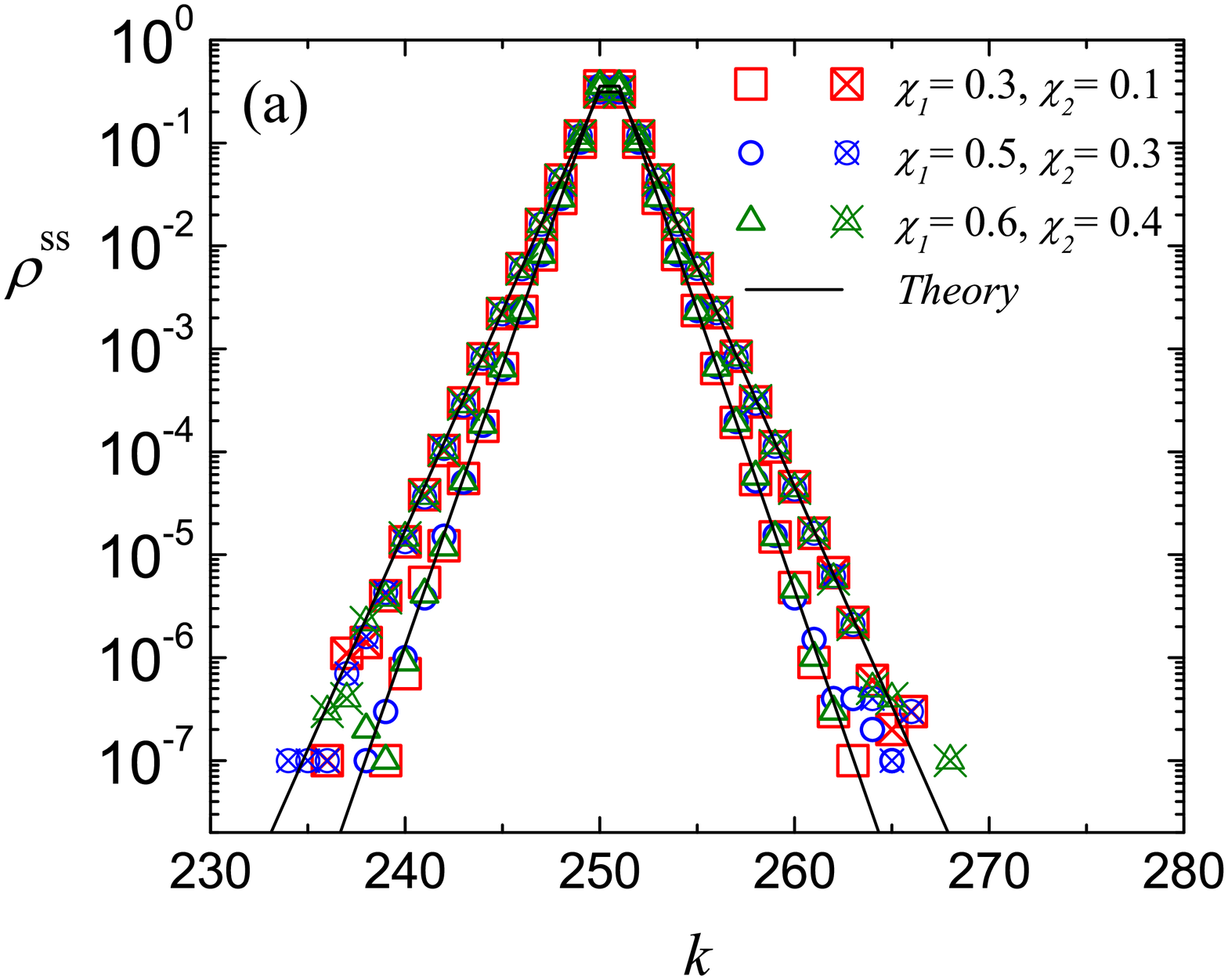}}\quad
    \subfigure{\includegraphics[width=3in]{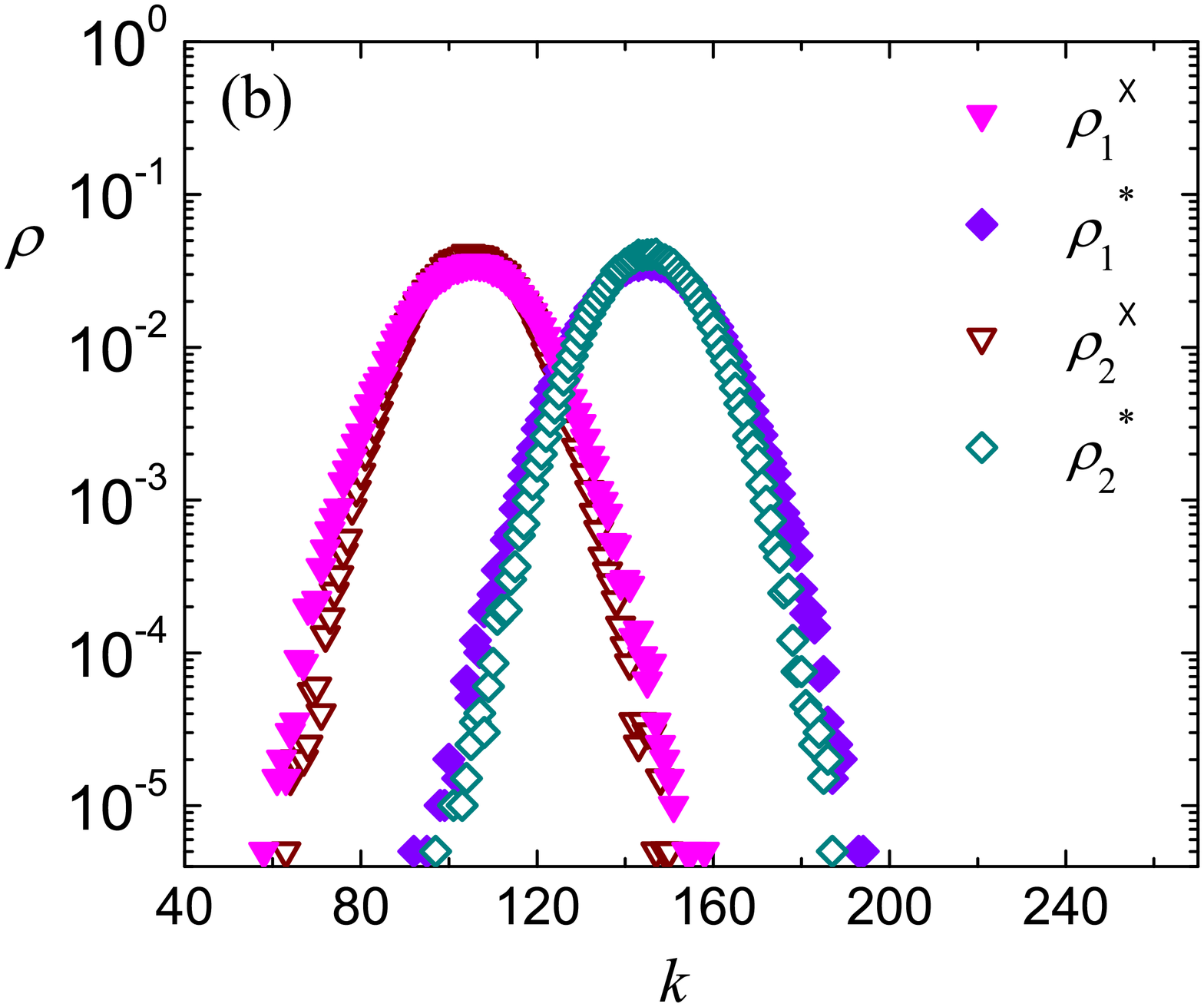}}
     }
\caption{(a) The markers without and with cross show simulation data for $\rho_1^{ss}$ and $\rho_2^{ss}$, respectively, with $N_1=N_2=N/2=1000$, $\kappa_1=\kappa_2=250$ and fixed $\epsilon=\chi_1-\chi_2=0.2$. The solid lines represent the analytic results. (b)Simulation results for internal and cross degree distributions for the system with $\chi_1=0.5$ and $\chi_2=0.3$ in (a). Solid diamonds and triangles represent $\rho_1^{\ast}$ and $\rho_1^{\times}$, and empty diamonds and triangles stand for $\rho_2^{\ast}$ and $\rho_2^{\times}$.}
\label{Nkappa}
\end{figure*}

Following our study of the homogeneous population, we begin with two \textit{%
identical} groups ($N_{1}=N_{2}=1000$ and $\kappa _{1}=\kappa _{2}=250$)
interacting via various $\chi _{\alpha }$'s. Using a similar scheme --
discarding the first $2K$ MCS and taking $10^{4}$ measurements
separated by $100$ MCS, we first consider the total degree distributions in
the steady state, $\rho _{a}^{ss}$. Not surprisingly, these are
indistinguishable from the $\rho ^{ss}$ above, namely, exponential distributions. A more
interesting scenario appears when the two groups differ only by their $\chi $%
's, so that $\varepsilon \equiv \chi _{1}-\chi _{2}\neq 0$. In particular,
our simulations show that the $\rho _{\alpha }^{ss}$ are still two-tailed exponentials, but
with $\varepsilon $-dependent tails. Fig.~3(a) illustrates this
effect, as we see that results of various $\chi $'s collapse into two sets.
Intuitively, such a difference can be attributed to `frustration' (in the
common psychological sense). If $\chi _{1}\gg \chi _{2}$, members of the
first group will make frequent attempts to `reach out' to those in the
second group. Since this behavior is not reciprocated, we may expect this
difference to be manifested in $\rho _{1,2}^{ss}$.

Whether we label the observed difference as `frustration' or not, the
significant message here is the following. Since the $1/2$ in (\ref{W})
accounts for the actions by all the other nodes, we must modify it to reflect the different contributions arising from \textit{inter}- \textit{vs} \textit{intra-}group nodes. Thus, when considering a node in, e.g., group $1$, the former
contribution is $\chi _{2}$ and the latter is $1-\chi _{1}$. The result of
such considerations is the equation%
\begin{equation}
\frac{\rho _{1}^{ss}(k)}{\rho _{1}^{ss}(k-1)}=\frac{w(k-1)+\left(
1-\varepsilon \right) /2}{1-w(k)+\left( 1-\varepsilon \right) /2}
\label{rho^ss+chi}
\end{equation}%
and a similar one for $\rho _{2}^{ss}$. These lead to 
\begin{equation}
\mu _{1}=ln \frac{3-\varepsilon }{1-\varepsilon };~~\mu _{2}=ln \frac{%
3+\varepsilon }{1+\varepsilon }
\end{equation}%
showing that the exponential decay rates deviate from $ln3$ in opposite directions. Remarkably, this
simple generalization of the argument advanced above provides a satisfactory
explanation, as illustrated by the black lines in Fig.~3(a).

Next, we turn to the separate distributions $\rho ^{\ast ,\times }$ (also in
the steady state, but we suppress the superscript $^{ss}$ for simplicity). Here we
encounter several surprises. The first is that these distributions appear
to be Gaussians, despite the total $\rho _{\alpha }^{ss}$ being two-tailed exponential!
To illustrate this finding, we provide four distributions $\rho _{1,2}^{\ast ,\times }$
in Fig.~3(b), for the case $\chi _{1}=0.5,\chi _{2}=0.3$. To resolve this
quandary, we turn to a better quantity for describing the connectivity
associated with a node in a system with two groups, namely, the \textit{%
joint }distribution $\Phi \left( k^{\ast },k^{\times }\right) $. (There
should be no confusion, as we dropped the subscript $\alpha $.) Representing
the probability that a node has $k^{\ast }$ intra-group links \textit{and} $%
k^{\times }$ inter-group links, it is related to the $\rho $'s by
projections 
\begin{eqnarray}
\rho \left( k\right) &=&\sum_{k^{\ast },k^{\times }}\delta \left( k^{\ast
}+k^{\times }-k\right) \Phi \left( k^{\ast },k^{\times }\right) \\
\rho ^{\ast }\left( k^{\ast }\right) &=&\sum_{k^{\times }}\Phi \left(
k^{\ast },k^{\times }\right) ; \quad \quad \rho ^{\times }\left( k^{\times }\right)
=\sum_{k^{\ast }}\Phi \left( k^{\ast },k^{\times }\right) .  \label{rho*,x}
\end{eqnarray}%
In the above case, we can describe $\Phi $ as a relatively sharp
`ridge,' situated along $k^{\ast }+k^{\times }\cong 250$ and descending
very steeply (exponentially) to the `valley floor.' As we move along this ridge,
the variations are more gentle, but as we venture further from the summit, $\Phi $ begins to descend as a Gaussian) . Thus, both of the simple
projections show the Gaussian profile, while the special projection reveals the
exponential distribution.

Examining these Gaussians closer, we see that both $\rho ^{\times }$'s have
means around $100$, while both $\rho ^{\ast }$'s are spread around $150$.
Further all are typically much broader, $O\left( 10\right) $, than the
two-tailed exponential distributions whose width is $O\left( 1\right) $. Further, it is remarkable that the $\rho
^{\times }$ and $\rho ^{\ast }$ data can be described, very roughly, by the
binomial distributions $\binom{250}{k}\left( 0.4\right) ^{k}\left(
0.6\right) ^{250-k}$ and $\binom{250}{k}\left( 0.6\right) ^{k}\left(
0.4\right) ^{250-k}$ respectively. This na\"{\i}ve picture comes from
partitioning $\kappa =250$ links into $100$ crosslinks (average of $\chi
_{1}\kappa _{1}$ and $\chi _{2}\kappa _{2}$, as a rough guess) and $150$
intra-group links. We should caution the reader that, despite repeated
reference to Gaussians and binomials, these distributions are of course, not
precisely so. To study their details, such as skewness and kurtosis, is
beyond the scope of this work. Instead, we will discuss a more serious
challenge in the next paragraph.

The second discovery is more intriguing, namely, the emergence of two very
different time scales. While it takes only $O\left( 10^{3}\right) $ MCS for
an initially empty network to `relax' into systems with $\left\langle
k\right\rangle \thicksim \kappa =250$, we find that $O\left( 10^{4}\right) $
MCS is sufficient for us to collect good data for $\rho _{\alpha }^{ss}$. We
will refer to this as the short time scale: $\tau _{short}$. During this
period, the distributions $\rho ^{\times ,\ast }$ appear to settle into the
Gaussians noted above. However, when examined at much later times (e.g., $%
O\left( 10^{5}\right) $ or $O\left( 10^{6}\right) $ MCS), the centers of
these Gaussians appear to \textit{wander} slowly (though their widths are
essentially unchanged). In other words, there is a much larger time scale,
after which the system finally relaxes into the true steady state. Denoted
by $\tau _{long}$, it appears to be much greater than {$O(100N)$
MCS. }Such behavior is observed even for the most symmetric case {$%
N_{1}=N_{2},\kappa _{1}=\kappa _{2},{\chi _{1}=}\chi {_{2}}=1/2$}! We will
return to this puzzle later. Here let us turn to other parameter choices for our two
interacting networks.

\subsubsection{Equal $N$'s and $\protect\chi $'s, but {${\protect\kappa _{1}.}%
\neq \protect\kappa {_{2}}$}}

\begin{figure*} 
  \centering
  \mbox{
    \subfigure{\includegraphics[width=3in]{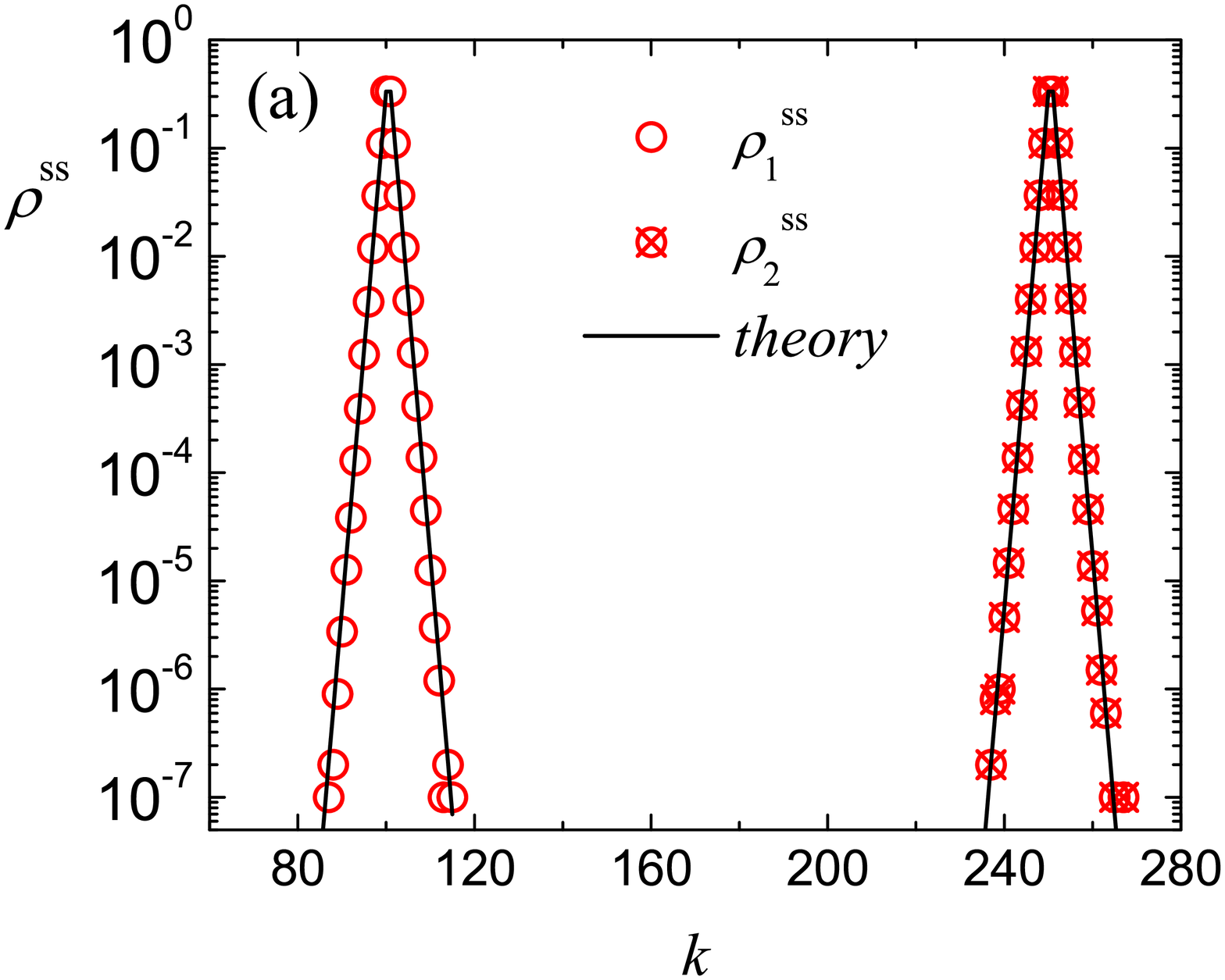}}\quad
    \subfigure{\includegraphics[width=3in]{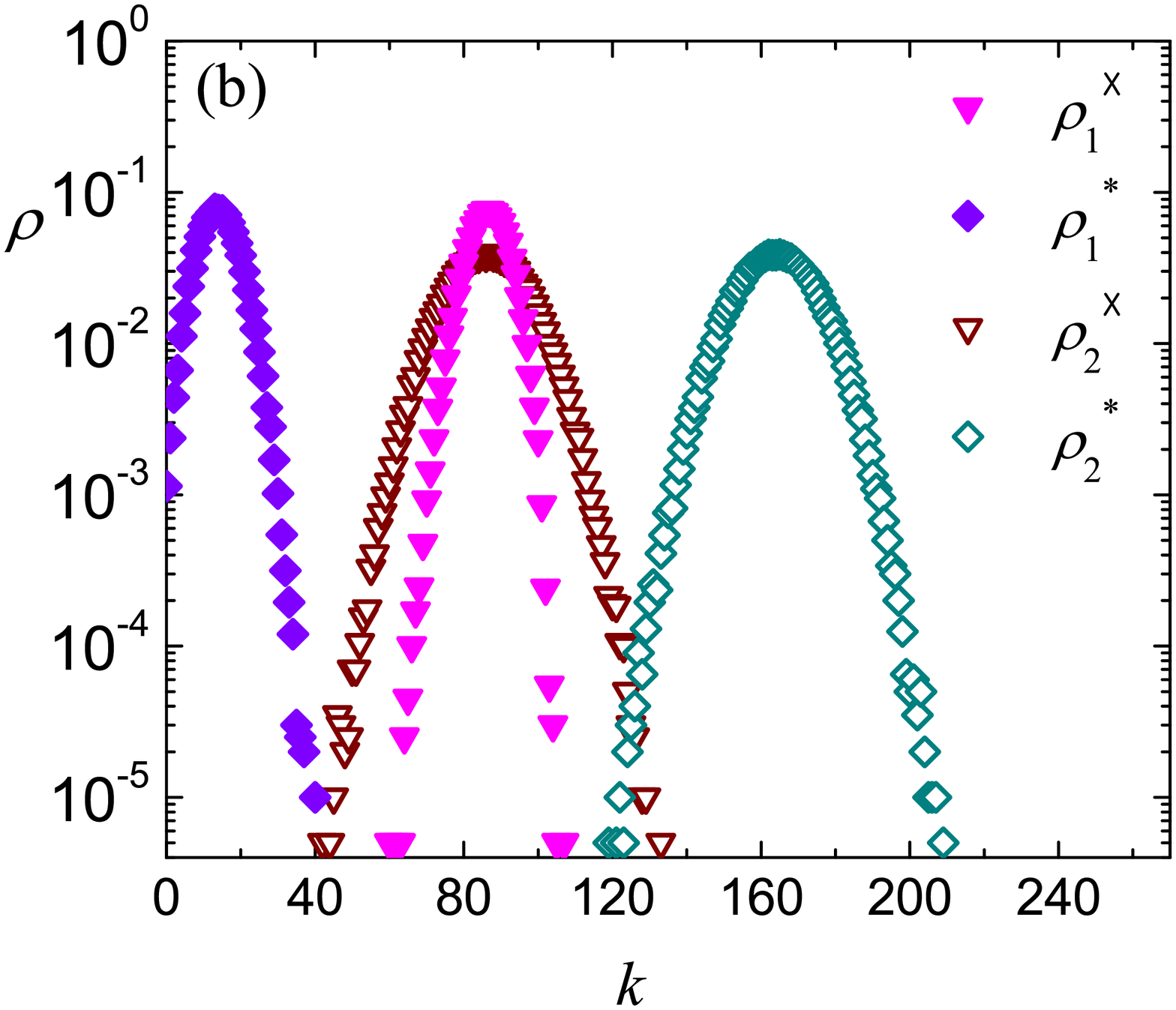}}
     }
\caption{(a) $\rho^{ss}$, (b) $\rho^{\star}$ and $\rho^{\times}$ for network one and network two, with parameters $N_1=N_2=N/2=1000$, $\kappa_1=100$, $\kappa_2=250$, and $\chi_1=\chi_2=0.5$. Solid squares and triangles represent $\rho_1^{\ast}$ and $\rho_1^{\times}$. Empty squares and triangles stand for $\rho_2^{\ast}$ and $\rho_2^{\times}$. }
\label{Nchi}
\end{figure*}

As indicated above, one main motivation for studying two groups is the
general perception that there are introverts and extroverts in our society.
Thus, we explore a simple initial step: all parameters being equal except the 
$\kappa $'s. We expect that such a system should display `frustration'
(again, in the psychological sense). The `introverts' are `frustrated' since the `extroverts' reach out to them, generating more links than the introverts prefer; in turn, the extroverts are also dissatisfied by seeing their links constantly cut by the introverts.
However, by exploring systems with only moderate differences, specifically, $%
\kappa _{1}=100$ and $\kappa _{2}=250$, we detect no conspicuous signs of
such frustration. Indeed, we find no new surprises here. Illustrated in Fig.~4 are degree distributions (obtained from short runs, $\tau _{short}$) for
the case with $N_{1}=N_{2}=1000$ and $\chi _{1}=\chi _{2}=0.5$. In Fig.~4(a),
we see the familiar two-tailed exponential distributions for the total degree
distributions -- with means at the two different $\kappa $'s and tails of $%
\mu \cong \ln 3$. The remaining distributions, shown in 
Fig.~4(b), are also
similar to the above, i.e., being essentially Gaussians. The most prominent
feature here is that their means and widths appear quite disparate. Of
course, we must have $N_{1}\left\langle k_{1}^{\times }\right\rangle
=N_{2}\left\langle k_{2}^{\times }\right\rangle $ since both equal the
average \textit{total }number of crosslinks in the systems. If we use the 
\textit{ad hoc} scheme above -- declaring the average crosslinks for each
node to be $\left( \chi _{1}\kappa _{1}+\chi _{2}\kappa _{2}\right) /2$, we
arrive at $87.5$. The complements are intra-group links, i.e., $12.5$ and $%
162.5$ here. As can be seen in the figure, these values are roughly
acceptable. The binomials, depicted above, give similarly rough portraits of
these $\rho $'s. The main challenge, as described earlier, remains to be the
understanding of the slow wandering of these $\rho $'s at much longer time
scales.

\subsubsection{Equal $\protect\kappa $'s and $\protect\chi $'s, but {${N_{1}}%
\neq N{_{2}}$}.}

\begin{figure*} 
  \centering
  \mbox{
    \subfigure{\includegraphics[width=3in]{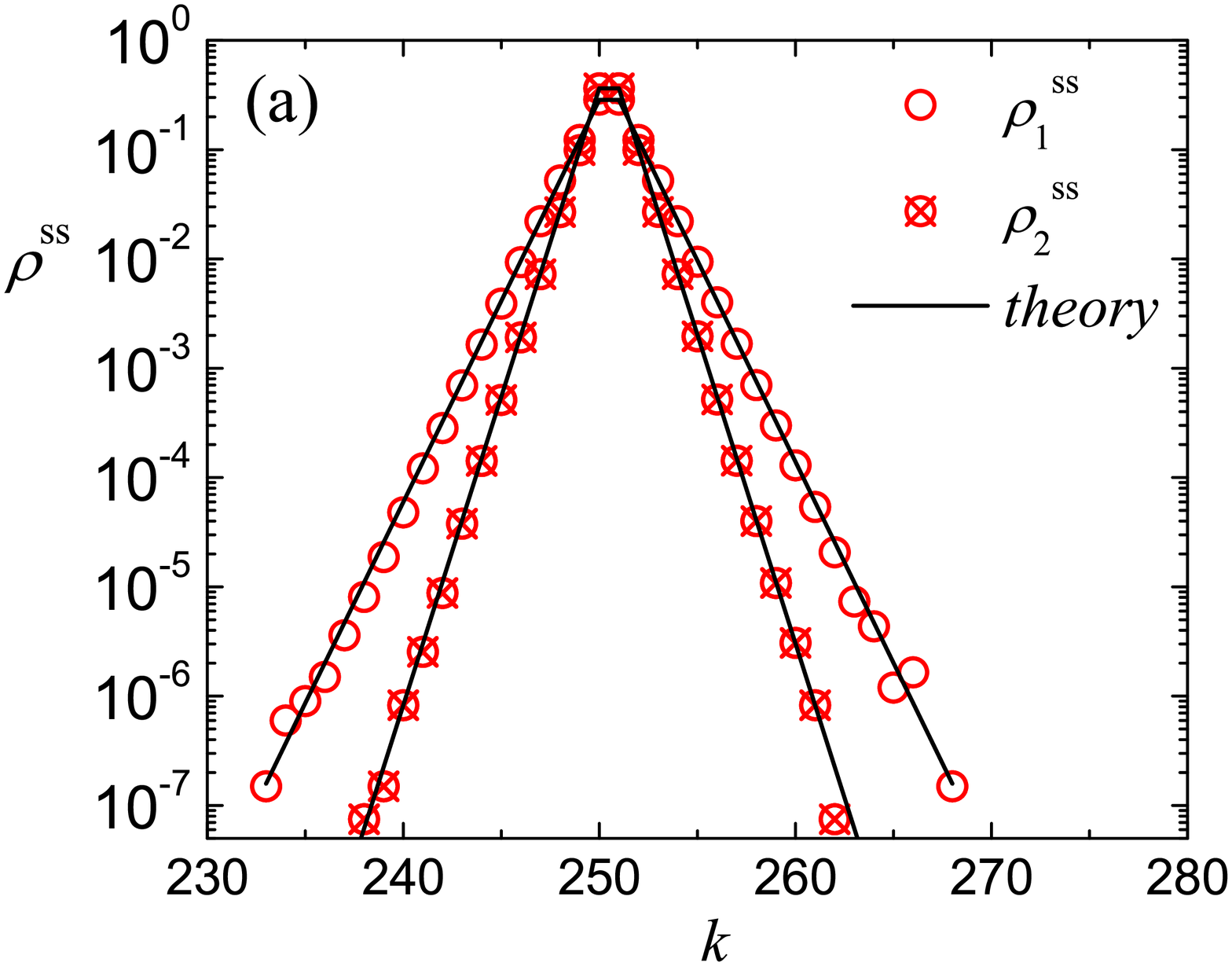}}\quad
    \subfigure{\includegraphics[width=3in]{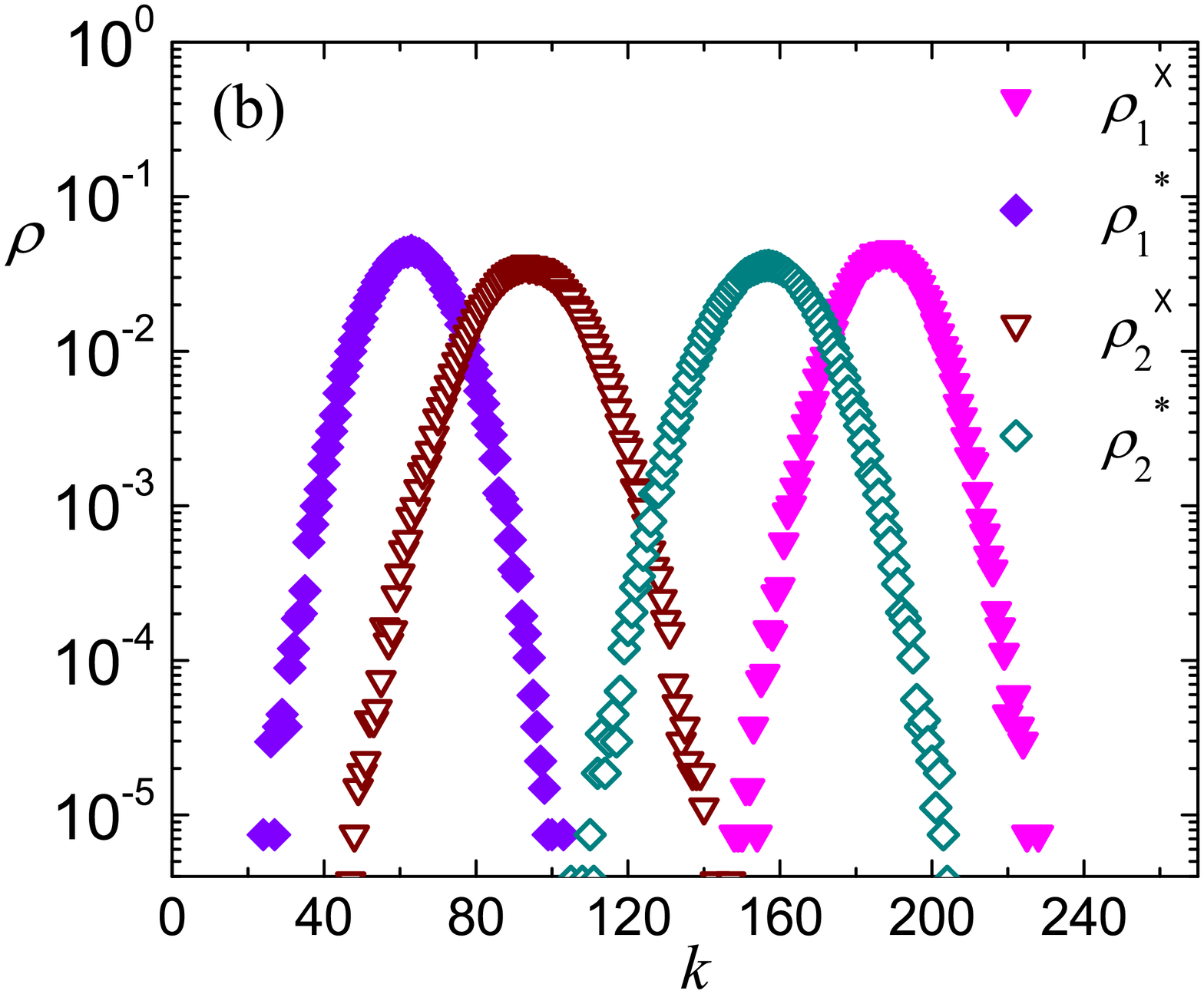}}
     }
\caption{(a) $\rho^{ss}$, (b) $\rho^{\ast}$ and $\rho^{\times}$ for network one and network two, with parameters $N_1=500$, $N_2=1000$, $\kappa_1=\kappa_2=250$, and $\chi_1=\chi_2=0.5$. Solid squares and triangles represent $\rho_1^{\ast}$ and $\rho_1^{\times}$. Empty squares and triangles stand for $\rho_2^{\ast}$ and $\rho_2^{\times}$. }
\label{chikappa}
\end{figure*}

To complete the skeletal picture, let us report some findings of systems
with unequal group sizes. In particular, we simulated a system with $N_{1}=500,N_{2}=1000$ ($\kappa _{1}=\kappa _{2}=250,\chi _{1}=\chi _{2}=0.5$), again for short periods of time ($\tau _{short}$). The various
degree distributions are shown in Fig.~5. Not surprisingly, the totals have
settled into two-tailed exponential distributions, though with different $\mu $'s. Since
the $\chi $'s are the same, the simple argument leading to 
Eq.~(\ref{rho^ss+chi}) is modified only by the different number of nodes in each group.
Specifically, if we again focus on a node in group $1$, then the (adding/cutting) actions of
the \textit{inter}-group nodes would be enhanced by a factor of $N_{2}/N_{1}$. If the $\chi $'s were also different, then the $1/2$ in Eq. (\ref{W}) would
become%
\begin{equation}
\frac{1}{2}\left\{ \frac{N_{2}}{N_{1}}\chi _{2}+\left( 1-\chi _{1}\right)
\right\} .  \label{Nchi}
\end{equation}%
In case {${\chi _{1}=}\chi {_{2}}=0.5$}, this factor becomes 
$N/4N_{1}$, so that we can use%
\begin{equation}
\frac{\rho _{\alpha }^{ss}(k)}{\rho _{\alpha }^{ss}(k-1)}=\frac{w(k-1)+%
N/4N_{\alpha }}{1-w(k)+N/4N_{\alpha }}  \label{rho^ss+N}
\end{equation}
to predict approximate distributions (black lines in Fig.~5(a)). Again, we
find very good agreement with the data here.

Meanwhile, for the separate distributions $\rho _{1,2}^{\ast ,\times }$, we
again observe Gaussians, though the $N_{1}\left\langle k_{1}^{\times
}\right\rangle =N_{2}\left\langle k_{2}^{\times }\right\rangle $ but the
means are now located at \textit{four} distinct values. As above, it is
possible to provide rough estimates for these results. Since $%
N_{1}\left\langle k_{1}^{\times }\right\rangle =N_{2}\left\langle
k_{2}^{\times }\right\rangle $, we are not surprised that $\rho _{1}^{\times
}$ peaks at $\thicksim 185$, a value twice that for $\rho _{2}^{\times }$, $%
\thicksim 90$. Arguably, we may regard these mean degrees as the result of
an effective $\chi $, i.e., $\left\langle k_{\alpha }^{\times }\right\rangle
=\kappa _{\alpha }\tilde{\chi}_{\alpha }$, giving us $\tilde{\chi}_{1}=2%
\tilde{\chi}_{2}$. If we further impose an \textit{ad hoc} assumption 
-- namely, that the average of these $\tilde{\chi}$'s should not change -- then we
arrive at $\tilde{\chi}_{1}=2/3$ and $\tilde{\chi}_{2}=1/3$. Remarkably,
such rough arguments differ only about $10\%$ from the simulation results.
While this approach cannot be taken as a good understanding of the phenomena
observed, it may provide a stepping stone towards a more reliable theory.

\subsection{Statistical properties of the total number of crosslinks, $X$.}
\begin{figure}[b]
\centering
\includegraphics[width=4in]{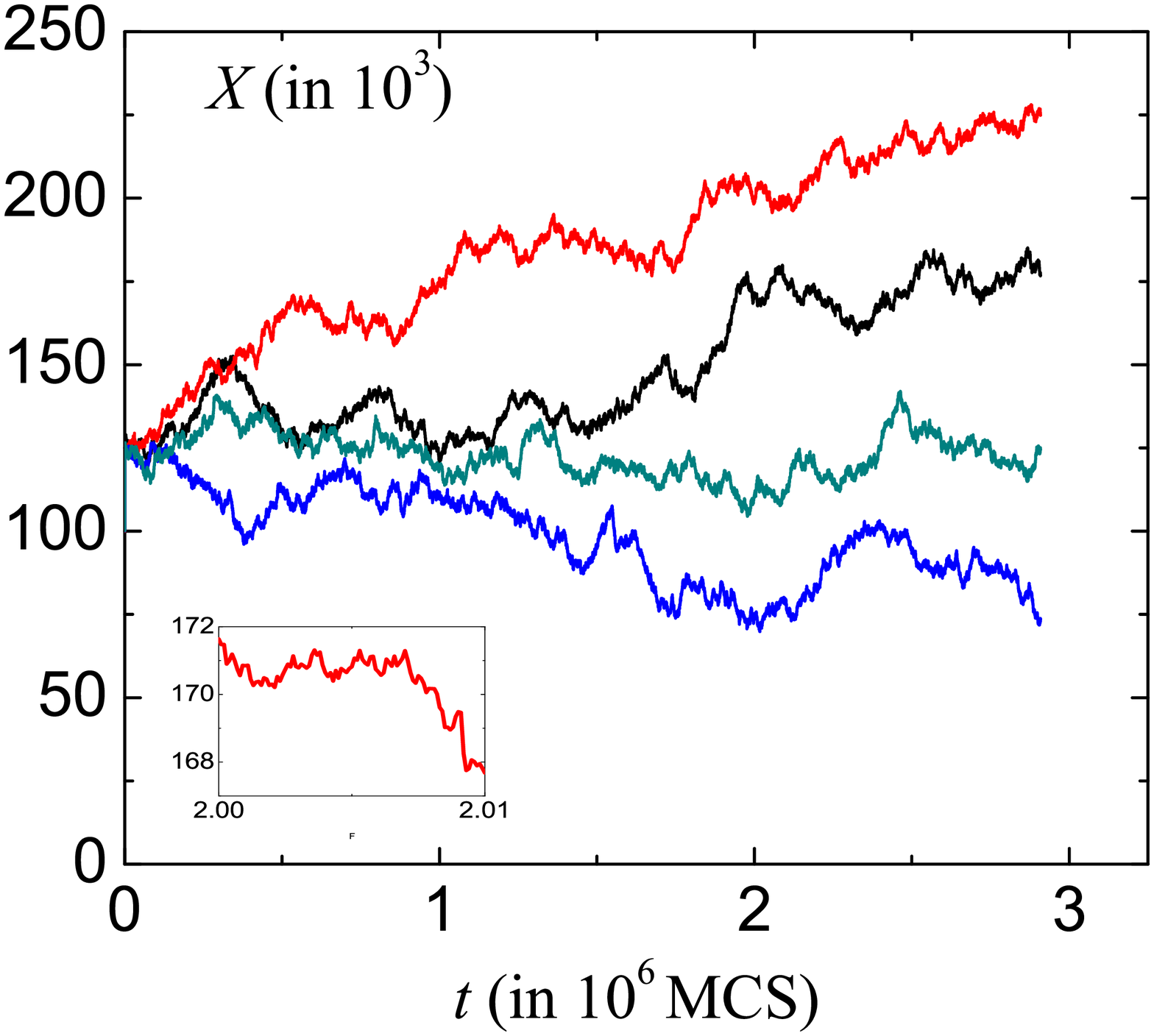}%
\caption{Four independent time traces $X(t)$ for a system with $N_1=N_2=N/2=1000$, $\kappa_1=\kappa_2=250$ and $\chi_1=\chi_2=0.5$. In the inset, we show a small section ($10^5$ MCS) of the red trace, to illustrate how little $X$ varies at this time scale. Note the scale for $X$ here spans just $4K$, compared to the $250K$ in the main figure.} 
\label{X}
\end{figure}
Though degree distributions are standard quantities for characterizing
networks, we have seen that, in a system with just two groups, additional
challenges arise when we consider distributions of different types of links.
The puzzles uncovered can be traced to, we believe, a single characteristic
of such systems, namely, $X$, the total number of links \textit{between} the
two groups. In particular, the slow wanderings of the means in $\rho
_{\alpha }^{\times }$ can be related to the slowly varying $X\left( t\right) 
$, while at shorter time scales, $\rho _{\alpha }^{\times }\left( k\right) $
is well described by a random distribution of $X$ among the $N_{\alpha }$
nodes. This subsection is devoted to a few initial steps towards the
understanding of the behavior of $X$.

To connect with the results from the last subsection and to emphasize the
challenge we face, we show the data associated with an apparently symmetric
system: {$N_{1}=N_{2}=1000$, $\kappa _{1}=\kappa _{2}=250$, ${\chi _{1}=}%
\chi {_{2}}=0.5$. These} parameters are chosen to be comparable to those studied in the previous subsections. In Fig.~\ref{X}, four runs of $X\left( t\right) $ over $3M$
MCS are not inconsistent with the traces of random walkers. Note
that, in \textit{all} cases, $X\left( 0\right) =0$ (we start with empty networks) and at
early times, $X\left( \tau _{short}\right) \thicksim 125K$, a number
consistent with the simple estimate {$N_{\alpha }\kappa _{\alpha }{%
\chi _{\alpha }}$}. Thereafter, $X$ wanders widely. Of course, this random
walk is bounded, by $0$ from below and $\thicksim{N_{\alpha }\kappa
_{\alpha }}$ (${=}250K$ here) from above. The latter is an estimate, assuming that every node has \textit{all} of its $O\left( \kappa \right) $
connections as crosslinks. From the figure, we see that these $4$ runs have not yet reached these boundaries. In other words, it would take $\tau
_{long}\gg 3M$ MCS for this system to finally settle in the true steady
state. Meanwhile, as the inset shows, $X$ is relatively constant within any
interval of $\tau _{short}\thicksim 10K$ MCS, while the fast mixing of
crosslinks between the nodes allows an individual $\rho _{\alpha }^{\times }$
to relax into an approximate Gaussian. Indeed, given $X$, these \textit{%
quasi-stationary} distributions fit quite well to another binomial, namely, $%
\binom{X}{k}\left( N_{\alpha }^{-1}\right) ^{k}\left( 1-N_{\alpha
}^{-1}\right) ^{X-k}$.

\begin{figure}
\centering
\includegraphics[width=4in]{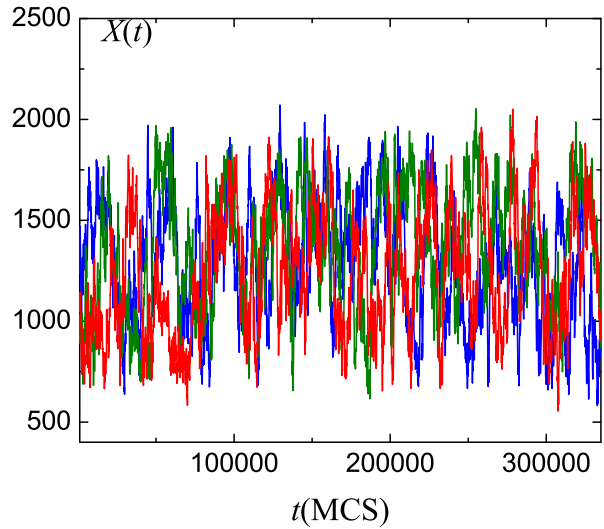}%
\caption{Three time traces of $X$ (red, green, and black), for a system with $N_1=N_2=N/2=100$, $\kappa_1=\kappa_2=25$ and $\chi_1=\chi_2=0.5$.} 
\label{crossRW-new}
\end{figure}

The next natural step is to probe deeper into the hypothesis that $X\left(
t\right) $ is indeed an unbiased random walk, between some `soft walls' $%
X_{\min }$ and $X_{\max }$ . But, to reach steady state after say, $10M$ MCS,
we must consider much smaller systems (along with smaller $\kappa $'s). For
example, with a system an order of magnitude smaller, {$N_{\alpha }=100$ and 
$\kappa _{\alpha }=25$}, we expect a random walk to traverse the full range
of $\lesssim 2500$ in about $\left( 2500\right) ^{2}$ steps. Since a step in 
$X$ will occur in just a few attempts, a good estimate of the traversal time
is $\left( 2500\right) ^{2}/200\thicksim 6000$ MCS. Thus, we can expect the
system to settle over runs of $\tau _{long}\thicksim 10^{7}$ MCS. When Monte
Carlo simulations are carried out (using the most symmetric case: {$%
N_{\alpha }=100,\kappa _{\alpha }=25,{\chi _{\alpha }}=0.5$}), these
expectations are indeed borne out. In Fig. \ref{crossRW-new}, we display
three short sections (each $10^{5}$ MCS long), obtained from partitioning a single long run ($10^{7}$ MCS). Note that $X\left( t\right) $ indeed traverses
the full range in each case.

\begin{figure}
\centering
\includegraphics[width=4in]{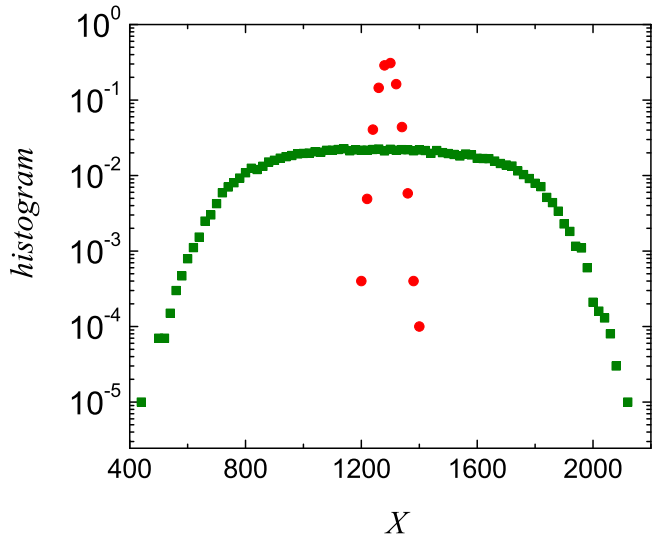}%
\caption{This figure shows the histogram of $X$ for two networks with $N_1=N_2=N/2=100$, $\kappa_1=\kappa_2=25$ and $\chi_1=\chi_2=0.5$ (olive squares), as well as the histogram of $X$ for a randomly labeled single network (red circles) with $N=200$ and $\kappa=25$.}
\label{histogram}
\end{figure}

With confidence that the system has reached steady state, we compile a
histogram, $P^{ss}\left( X\right) $, from this trace and show the result
(olive squares) in Fig.~\ref{histogram}. Note that this distribution is
relatively flat, around the mean of approximately $1250$ (i.e., {$N_{\alpha }\kappa
_{\alpha }{\chi _{\alpha }}$}) with soft cutoffs at both ends. Such a wide
and fairly flat `plateau' in $P^{ss}\left( X\right) $ is consistent with the
idea that $X\left( t\right) $ executes a simple random walk between two
soft walls, located approximately at $X_{\min }\thicksim 600$ and $X_{\max }\thicksim 1900$.

In the next paper, we will provide other measures which strengthen our
hypothesis. Here, let us end with addressing a natural question: Is there
any difference between our `most symmetric' case ({$N_{\alpha }=100,\kappa
_{\alpha }=25,{\chi _{\alpha }}=0.5$}) and a single homogeneous network of $%
200$ nodes with $\kappa =25$? In particular, what can be expected if we
arbitrarily label half of the latter nodes as `red' and the rest `blue,' and
compile a histogram for the total number of red-blue links in the system?
Simulations show a remarkably different picture. Illustrated with solid
red circles in Fig.~\ref{histogram} this distribution is much sharper than $P^{ss}\left( X\right) $ and well described by a Gaussian, with mean close to 
$1275$ and standard deviation $\sigma \thicksim 25$. The value of the mean
is not surprising, especially if we recall that, in our simulations, the
effective $\kappa $ is $25.5$. As for $\sigma $, it is precisely the value
of the most na\"{\i}ve expectation, from applying the central limit theorem
to adding $100$ random variables distributed according to $\left. \binom{25}{%
k}\right/ 2^{25}$. It is remarkable how two models which
appear to be so similar exhibit such drastically different behavior.
In particular, by modeling interactions between two identical groups with a single parameter, $\chi $, we encounter counter-intuitive phenomena.

\section{SUMMARY AND OUTLOOK}

In this paper, we introduced the idea of a stochastically evolving network
with preferred degrees. The key feature of our models is that each node can
add or cut one of its links, depending on whether it finds itself with too
few or too many (compared to some built-in `preferred' number of) contacts.
To establish a baseline, we focus first on a single homogeneous isolated
network, in which every node `prefers' degree $\kappa $. Specifically, when
chosen, a node (with degree $k$) will create or destroy a link with
probability $w_{\pm }(k;\kappa )$. For simplicity, we only studied models
with $w_{-}=1-w_{+}$, while modeling how `intolerant' an individual is, when
it finds $k\neq \kappa $, by the expression (\ref{w}). We showed  that even in such simple models, the dynamics violates detailed
balance, so that the long-time limit is a non-equilibrium steady state. With
generally unknown probability distributions and non-trivial probability
currents, an exact and analytic approach is all but impossible. Instead, we
explore the statistical properties by Monte Carlo simulations and a variety
of mean-field approaches. Simulating mainly the most rigid ($\beta =\infty $) population, we discovered that an initially empty network of $1000$ nodes
(with $\kappa =250$) settles into a steady state quite quickly ($\thicksim
10^{4}$ MCS). The degree distribution is a double
exponential, around $\kappa $: $\rho^{ss} (k)\propto e^{-\mu \left\vert k-\kappa
\right\vert }$. A simple mean-field argument, in the context of an
approximate master equation, leads to $\mu =\ln 3$, which agrees well with
data. For more flexible populations, $\rho^{ss} (k)$ is Gaussian-like around $%
\kappa $ and, for large $\left\vert k-\kappa \right\vert $, crosses over to
the exponentials above. Our mean-field theory can be generalized appropriately and
provides similarly good agreement with the simulation results. Of course,
the system will not display this type of behavior for extreme values of $N$, 
$\kappa $ and $\beta $ (e.g., near zero) and we believe the theory will
break down in those limits. Nevertheless, for generic points in parameter
space, we are confident that the main features of this adaptive network are
not difficult to understand, both intuitively and quantitatively.

We then introduced a second preferred degree network and coupled it to the
first, through $\chi $, the probability that a node adds or cuts a crosslink
(between the networks). With two networks, the parameter space is already so
large that a completely systematic study is beyond our scope. We focused on
three cases where the two networks differ by only one of the three
parameters ($N,\kappa ,\chi $). Seemingly a simple extension of the
homogeneous case, this model provides a rather wide range of interesting
results, from the mundane and comprehensible to the surprising and puzzling.
The total degree distribution of each network is not seriously affected by
the interaction and can be reasonably well explained by extending the
approximation scheme for the single network case. By contrast, serious
challenges emerge when we consider the more detailed distributions: $\rho
^{\ast }$ and $\rho ^{\times }$, associated with intra- and inter-group
degrees, $k^{\ast }$ and $k^{\times }$, respectively. Though both \textit{%
total} $\rho $'s remain two-tailed exponentials, all these new distributions are roughly
Gaussians, with means and widths that are yet to be clearly understood. More
importantly, we studied a global quantity which is suitable for
characterizing the inter-network interactions, namely, the total number of
crosslinks in the system, $X$. Remarkably, it displays a very slow dynamics,
as well as extensive fluctuations. For example, even after $10^{6}$ MCS, all
of our runs for $X$ (in the `most symmetric' case of $N_{\alpha
}=1000,\kappa _{\alpha }=250,\chi _{\alpha }=0.5$) exhibit what appears to be
an unbounded, unbiased random walk!\ By lowering the system parameters to $%
N_{\alpha }=100,\kappa _{\alpha }=25$ and running up to $10^{7}$ MCS, we
finally observed a stationary distribution for $X$. Being almost flat and broad, this $P^{ss}\left(
X\right) $ is also not well understood yet.

These initial findings provide us with first steps in this study of interacting networks. The next steps will be
presented in the next three papers of this series. Let us provide a preview of the rest of the series. In a second paper, we will present a
more systematic study of the statistical properties of $X$ as a function of the
parameters of the two preferred degree networks: $\left( N_{\alpha },\kappa
_{\alpha },\chi _{\alpha }\right) $, $\alpha =1,2$. Since the underlying
dynamics does not obey detailed balance, an explicit expression for the the microscopic stationary distribution through equilibrium statistical mechanics is not possible. As a result, we rely mostly on Monte Carlo simulations. A certain level of theoretical understanding can be obtained from various
approximation schemes for a master equation governing the evolution of 
$P\left( X,t\right) $. In the third paper, we will consider an extreme limit: 
$\kappa _{1}=0$ and $\kappa _{2}\rightarrow \infty $. We coin the name `XIE' model for this case of e\b{x}treme \b{i}ntroverts and \b{e}xtroverts, in
which every introvert prefers zero contacts (and only cuts links) and every
extrovert prefers as many friends as possible (and always adds links). This
limit is interesting for several reasons. The only relevant parameters left
are $N_{1}$ and $N_{2}$. Meanwhile, detailed balance is
restored in this limit and so, an explicit microscopic stationary
distribution of the system can be obtained. Nevertheless, 
$P^{ss}\left(X\right) $ cannot be computed analytically, though a mean field
approach seems to be quite adequate for predicting its key features. Most
surprisingly, there is an extraordinary transition in the system ($\chi
_{1}=\chi _{2}$), as the ratio $N_{1}/N_{2}$ is varied through unity \cite
{LiuSchmittmannZia12}. Further, using a self-consistent mean field approximation, we are
able to predict (with no fit parameters!) $\rho _{\alpha }^{ss}\left(
k\right) $, except for the case $N_{1}=N_{2}$. In the last paper, we will
present results for models involving several other forms of interaction.
Perhaps more realistic, these will include letting an individual have two $%
\kappa $'s, to differentiate actions taken with inter- and intra-network
contacts. Clearly, our primary focus for this series rests on  the statistical properties
of systems in \textit{steady} states. The full time-dependent behavior of dynamic networks,
clearly much richer and more complex, will be considered in the future.

We conclude with a few comments on how to extend our model to more realistic cases. First, there is typically a full
spectrum of `preferences' in every society, and so one should really consider a set, $\left\{ \kappa _{i}\right\} $. Second, in our model here, every
individual can connect with every other one, which clearly fails to capture the more complex structures of a real society, from simple spatial proximities to social status and subtle ethnic divides, etc. Third, we should explore more realistic models
of real phenomena, where nodes (individuals) are endowed with dynamic degrees of freedom, e.g., opinions, wealth, or health. These degrees of freedom in turn determine the connections (links) between individuals in a society, leading to a fully co-evolving model of node and link dynamics.

Beyond social networks, a worthy goal would be
the understanding of the interactions between dramatically different
networks, such as those listed in the Introduction. Clearly, achieving such
a goal would have significant and long-term impact on both network science
and the welfare of our species.

\ack

We thank K. Bassler, H. Kim, M. Pleimling, T. Platini, L.B. Shaw and Z. Toroczkai for illuminating
discussions. This research is supported in part by ICTAS, Virginia Tech, and NSF grant DMR-1244666.

\appendix

\section{Stochastic dynamics of a single network: Exact master equation
and violation of detailed balance}

A single network can be described by an $N\times N$ adjacency matrix $\mathbb{%
A}$ (symmetric in our case, as the links are undirected), where the elements 
$A_{ij}=0$ $(1)$ indicate the absence (presence) of the link between nodes $i$
and $j$. Since self-loops are not allowed, $A_{ii}=0$ for all $i\in
\lbrack 1,N]$. A complete analytical description of the stochastic evolution
of our model is provided by $\mathcal{P}(\mathbb{A},t~|\mathbb{A}_{0},0)$,
which is the probability of finding configuration $\mathbb{A}$ at time $t$,
given an initial configuration $\mathbb{A}_{0}$. Since our focus is on a
dynamics without memory, i.e., a Markov process, we can write down the
discrete master equation for $\mathcal{P}$ as follows. The change over one
attempt, $\mathcal{P}(\mathbb{A},t+1)-\mathcal{P}(\mathbb{A},t)$ is 
\begin{equation}
\sum_{\{\mathbb{A}^{\prime }\}}[R(\mathbb{A},\mathbb{A}^{\prime })\mathcal{P}%
(\mathbb{A}^{\prime },t)-R(\mathbb{A}^{\prime },\mathbb{A})\mathcal{P}(%
\mathbb{A},t)]  \label{ME}
\end{equation}%
where $R(\mathbb{A},\mathbb{A}^{\prime })$ is the rate for configuration $%
\mathbb{A}^{\prime }$ to change to $\mathbb{A}$. Note that, since each $%
\mathbb{A}$ has $\mathcal{L}\equiv N\left( N-1\right) /2$ elements, the
configuration space in which $\mathcal{P}(\mathbb{A},t)$ evolves consists of
the $2^{\mathcal{L}}$ vertices of a unit cube in $\mathcal{L}$-dimensional
space. In this setting, each attempt is seen to be just a step from one
vertex to another along an edge of this cube.

Explicitly, $R$ consists of a sum over terms, each corresponding to an
attempt at changing the state of a link. We begin with the probability to
choose a particular node, $i$: $1/N$. Next, we need its degree, $k_{i}$,
which is obtained by summing up all elements along, say, the row $i$ : $%
k_{i}=\sum_{j}A_{ij}$. From here, we attempt to add a link with probability $%
w\left( k_{i}\right) $, or cut with probability $1-w\left(
k_{i}\right) $. Consider first a cutting action, which can occur for one of
the $k_{i}$ existing links, so that the total probability for, say, $A_{ij}$
to change from $1$ to $0$ (by node $i$) is $\left[ 1-w\left( k_{i}\right) %
\right] /[Nk_{i}].$\ Meanwhile, none of the other links changes in this
attempt. Thus, the term describing this action is%
\begin{equation}
\Delta \frac{1-w\left( k_{i}\right) }{Nk_{i}}\left( 1-A_{ij}^{\prime
}\right) A_{ij}
\end{equation}%
where%
\begin{equation}
\Delta \equiv \Pi _{k\ell \neq ij}\delta \left( A_{k\ell }^{\prime
},A_{k\ell }\right)
\end{equation}%
A similar term can be written to describe the adding action. All together,
we have 
\begin{equation}
R(\mathbb{A},\mathbb{A}^{\prime })=\sum\limits_{i}\frac{\Delta }{N}%
\sum\limits_{j\neq i}\left[ \frac{1-w\left( k_{i}\right) }{k_{i}}\left(
1-A_{ij}^{\prime }\right) A_{ij}+\frac{w\left( k_{i}\right) }{N-1-k_{i}}%
A_{ij}^{\prime }\left( 1-A_{ij}\right) \right] .
\end{equation}

Once the rates are known explicitly, we can check if they satisfy detailed
balance or not. The Kolmogorov criterion \cite{Kolmogorov} states that a set
of $R$'s satisfies detailed balance if and only if the product of $R$'s around 
\textit{any} closed loop in configuation space is equal to that around the
reversed loop. In our case, all loops can be regarded as sums over
\textquotedblleft elementary closed loops,\textquotedblright\ i.e., ones
which goes around a plaquette (or face) on our $\mathcal{L}$-cube. Thus, we
only need to focus on such elementary loops. Clearly, such a loop involves
two links, e.g., by adding two links from a given $\mathbb{A}$, followed by
cutting them to return to the original $\mathbb{A}$. As a specific
example, suppose we start with an $\mathbb{A}$ which has neither an $ij$
link nor an $im$ one. Then the sequence%
\begin{equation}
\binom{A_{ij}}{A_{im}}=\binom{0}{0}\rightarrow \binom{1}{0}\rightarrow 
\binom{1}{1}\rightarrow \binom{0}{1}\rightarrow \binom{0}{0}
\end{equation}%
denotes adding these two and cutting them, while the rest of $\mathbb{A}$ is
left unchanged. Apart from an overall factor of $N^{-4}$, the
product of the $R$'s associated with this loop is 
\begin{equation}
\begin{split}
& \left( \frac{w(k_{i})}{N-1-k_{i}}+\frac{w(k_{j})}{N-1-k_{j}}\right) \left( 
\frac{w(k_{i}+1)}{N-1-(k_{i}+1)}+\frac{w(k_{m})}{N-1-k_{m}}\right) \times  \\
& \left( \frac{1-w(k_{i}+2)}{k_{i}+2}+\frac{1-w(k_{j}+1)}{k_{j}+1}\right)
\left( \frac{1-w(k_{i}+1)}{k_{i}+1}+\frac{1-w(k_{m}+1)}{k_{m}+1}\right) 
\end{split}
\label{loop}
\end{equation}%
Now, the reversed loop can be denoted as%
\begin{equation}
\binom{A_{ij}}{A_{im}}=\binom{0}{0}\rightarrow \binom{0}{1}\rightarrow 
\binom{1}{1}\rightarrow \binom{1}{0}\rightarrow \binom{0}{0}
\end{equation}%
associated with the product
\begin{equation}
\begin{split}
& \left( \frac{w(k_{i})}{N-1-k_{i}}+\frac{w(k_{m})}{N-1-k_{m}}\right) \left( 
\frac{w(k_{i}+1)}{N-1-(k_{i}+1)}+\frac{w(k_{j})}{N-1-k_{j}}\right) \times  \\
& \left( \frac{1-w(k_{i}+2)}{k_{i}+2}+\frac{1-w(k_{m}+1)}{k_{m}+1}\right)
\left( \frac{1-w(k_{i}+1)}{k_{i}+1}+\frac{1-w(k_{j}+1)}{k_{j}+1}\right) 
\end{split}
\label{reloop}
\end{equation}

Of course, we can find the difference explicitly and verify that it does not
vanish in general. To appreciate this fact more easily, note that, e.g., the
factor $w(k_{i})w(k_{m})$ appears in (\ref{loop}) but not in (\ref{reloop}).
From these considerations, we conclude that detailed balance is violated
here.

We should re-emphasize the following. In our case, the products of $R$'s
around many elementary loops are the same as those of the reversed loops
(e.g., two links involving 4 different vertices). However, detailed balance
is satisfied only if \textit{all} loops are `reversible.' So, showing just
one `failed loop'\ is sufficient for us to conclude that detailed balance is
violated, the consequences of which are quite serious (see, e.g., \cite
{ZS2007} for further details.).

\section{Approximation schemes for the transition rates, $W$}

We provide the simple arguments used for the transitions rates appearing in
Eqns.~(\ref{ME-rho}). These lead to slightly more sophisticated versions of
the expressions (\ref{W},\ref{rho^ss},\ref{rho^ss+chi},\ref{rho^ss+N}) in
the main text.

First, we consider the single network case and argue as follows to obtain a
simple expression for $W\left[ k-1,k\right] $, the probability that a node
with degree $k$ will lose one of its links. We focus on a particular node ($l$)
with degree $k_{l}$. In each attempt, the probability for the node itself to
be chosen is just $1/N$ and then for it to cut a link is $1-w(k_{l})$. In
addition, node $l$ can lose a link if one of the other $k_{l}$ nodes
connected to it (say node $m$) chooses to cut a link, \textit{and} to cut 
\textit{the} link to node $l$ (i.e., the $ml$ link here). Now, the probability
is $k_{l}/N$ for one of these nodes to be chosen. Assuming all nodes are
equally likely to have too many or too few links, we approximate the
probability for cutting to be $1/2$. Finally, if $m$ were chosen, then the
probability it cuts the $ml$ link is $1/k_{m}$, which we replace by $%
1/\kappa $, by invoking a mean-field approximation. Combining these factors,
the chance that node $l$ will lose a link due to the action of others is $\left(
k_{l}/N\right) \left( 1/2\kappa \right) $. Thus, we have 
\begin{equation}
W\left[ k_{l}-1,k_{l}\right] \cong \frac{1}{N}\left\{ 1-w(k_{l})+\frac{k_{l}%
}{2\kappa }\right\} .  \label{Wa}
\end{equation}%
A further approximation assumes $k_{l}\cong \kappa $ and we arrive at (\ref%
{W}). Clearly, a similar argument leads to the probability for adding links, 
$W\left[ k_{l},k_{l}-1\right] $, yielding a slightly different version of (%
\ref{rho^ss}). In the specific cases we studied, these two versions are so
similar that both predictions are consistent with the simulation data. If $%
\kappa $ were $O\left( 1\right) $, then we can expect more discernable
differences. Investigations along such lines remain to be undertaken.

Turning to systems with two populations, let us first consider those with
equal $N$'s and $\kappa $'s but {${\chi _{1}}\neq \chi {_{2}}$}. Let us focus on a
node $l$ in group $1$, so that the probability for it to be chosen is $1/%
N$ and for it to cut is again $1-w(k_{l})$. The new aspect here is
that there are two groups of nodes which may be connected to $l$, corresponding to a total of $%
k_{l}^{\ast }+k_{l}^{\times }$ links. Each of these has some probability that it
will cut its link to $l$. The chance of choosing from the $k_{l}^{\ast }$
(intra-group) nodes is $k_{l}^{\ast }/N$. As before, we assume
that $1/2$ is the probability such a node ($m$) will cut. Now, the novel
feature is that with probability $\left( 1-\chi _{1}\right) $ it  will cut an
intra-group link while $\thicksim 1/\left\langle k_{m}^{\ast }\right\rangle $
is the chance it will cut the $ml$ link. If we make the further
approximation $k_{l}^{\ast }\cong \left\langle k_{l}^{\ast
}\right\rangle =\left\langle k_{m}^{\ast }\right\rangle $ (since both are in
group $1$), then these considerations lead to $\left( 1-\chi _{1}\right) /2%
N$. A similar argument for the actions of a node in group $2$
leads to $\chi _{2}/2N$, so that we have 
\begin{equation}
W\left[ k_{l}-1,k_{l}\right] \cong \frac{1}{N}\left\{ 1-w(k_{l})+%
\frac{\left( 1-\chi _{1}\right) +\chi _{2}}{2}\right\} .
\end{equation}%
Combining a similar argument for $W\left[ k_{l},k_{l}-1\right] $, we arrive
at (\ref{rho^ss+chi}). A pattern for such considerations begins to emerge, so
that expressions such as (\ref{Nchi}) and (\ref{rho^ss+chi}) can be easily derived.

However, we should point out that it is much more unreliable to develop arguments like these for the case of {${\kappa _{1}}\neq \kappa {_{2}}$},
which is the reason for using \textit{ad hoc} schemes such as
\textquotedblleft average of $\chi _{1}\kappa _{1}$ and $\chi _{2}\kappa _{2}
$.\textquotedblright

\end{document}